# Jamming modulates coalescence dynamics of shear-thickening colloidal droplets

M. V. R. Sudheer,[1,#] Sarath Chandra Varma,[2,#] Aloke Kumar,[2] Udita U. Ghosh[1,*]

[1]*Department of Chemical Engineering and Technology, Indian Institute of Technology (BHU), Varanasi, 221005, India.*

[2]*Department of Mechanical Engineering, Indian Institute of Science, Bangalore, 5610012, India.*

## Abstract

Recent investigations into coalescence dynamics of complex fluid droplets revealed the existence of *sub-Newtonian* behavior for polymeric fluids (elastic and shear thinning). We hypothesize that such delayed coalescence or *sub-Newtonian* coalescence dynamics may be extended to the general class of shear thickening fluids. To investigate this, droplets of aqueous corn-starch suspensions were chosen and its coalescence in sessile-pendant configuration was probed by high-speed real-time imaging. Temporal evolution of the neck (growth) during coalescence was quantified as a function of suspended particle weight fraction, $\phi_w$. The necking behavior was found to evolve as the power-law relation, $R = at^b$ where $R$ is neck radius, with exponent, $b \leq 0.5$, implying it is a subset of the generic sub-Newtonian coalescence. Second, significant delay in the coalescence dynamics is observed for particle fractions beyond the jamming fraction, $\phi_w > \phi_J \geq 0.35$. Our proposed theoretical model captures this delay implicitly through altered suspension viscosity stemming from increased particle content.

Keywords : Droplet coalescence, Colloidal suspensions, Particle clusters, Neck evolution.

[#]equal contribution
[*]corresponding author
Email: udita.che@iitbhu.ac.in

# Graphical abstract

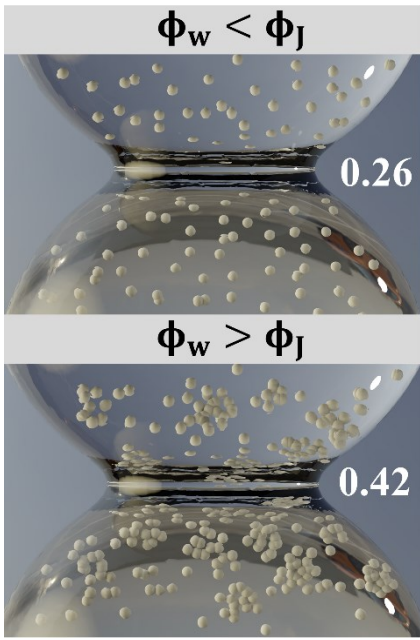
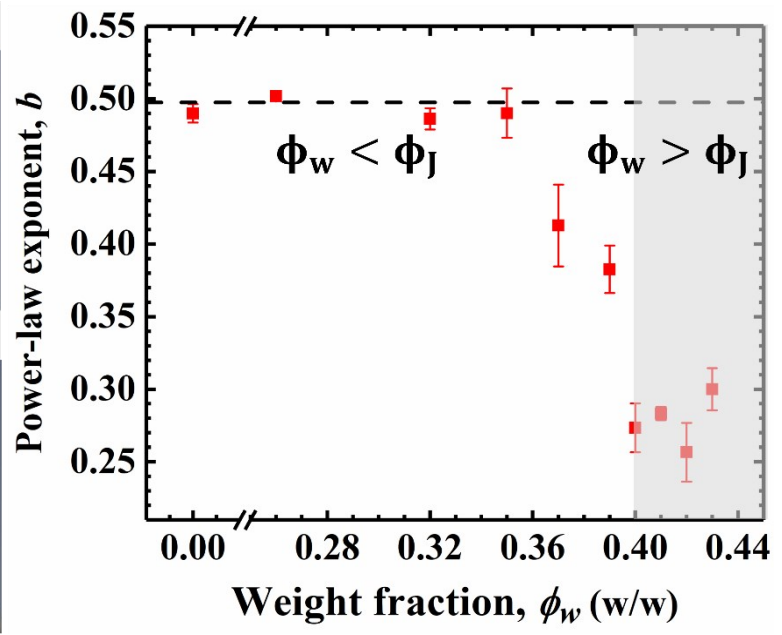

# 1. INTRODUCTION

Coalescence is the spontaneous merging of individual fluid droplets to form a single entity [1,2]. This process involves the emergence and growth of a liquid bridge between the coalescing droplets, called the neck. Temporal evolution of the neck can be uniquely captured by a single parameter, neck radius, $R$ known to evolve as, $R = at^b$ where, $a$ and $b$ are the pre-factor and power-law exponent respectively [3,4]. Magnitude of this power law exponent, $b$ is determined by the balance between the driving capillary force and the opposing inertia-viscous forces. There are three possible geometric configurations of coalescence – sessile-sessile, sessile-pendant and pendant-pendant. The present study is restricted to sessile-pendant configuration, where the exponent, $b$ is reported to be universal for coalescing droplets of Newtonian fluids [5,6]. The numerical values of the exponent are 0.5 and 1 in the inertia-dominated and viscous-dominated regimes resp. [5,6]. The parameter, $b$ singularly captures the alterations in the coalescence dynamics arising from changes in fluid properties, miscibility and ambient conditions. Recently, polymeric fluids droplet coalescence has been shown to have slower coalescence ($b < 0.5$) than the corresponding Newtonian coalescence [7–10]. This sluggish response has been attributed to restoring elastic force inherent to macromolecular relaxation processes in polymers. A gradual decrease in $b$ was also observed with an increase in the concentration of polymeric solute in the solution [8,10]. Thus, the existing theoretical framework saw the introduction of a new regime i.e., the elasticity-dominated regime [8,10] in case of polymeric fluids. It is therefore evident that the signature of the coalescence phenomenon, $b$ is specific to the class of fluids in question. A new term '*sub-Newtonian coalescence*' [7] has been coined for the class of fluids that exhibit coalescence dynamics with $b < 0.5$. So, the question that arises here is, is *sub-Newtonian* coalescence a characteristic of the restoring elastic force in the fluid? A recent theoretical investigation presents alternative results where coalescing droplets of shear thinning inelastic fluid in sessile-sessile configuration [11] exhibited *sub-Newtonian* coalescence. Similar effect has also been observed in coarse grained molecular scale simulations of coalescing droplets laden with surfactants. Introduction of surfactants slowed bridge growth and this deceleration has been attributed to formation of aggregates in droplet bulk that hindering fluid flow into the bridge region coupled with reduction in surface tension that delays initiation of bridge between the droplets [12]. Thus, it is evident that *sub-Newtonian* coalescence or delayed coalescence has been predicted to occur for different classes of complex fluid although the mechanism of occurrence is still unclear and fluid class-specific. In this study, therefore we choose a well-studied and characterised sub-class of complex fluids that is also microstructurally different from polymeric fluids and surfactant laden droplets – colloidal suspensions. The fundamental difference in their microstructure [13] arises from presence of short-range inter-particle forces, wherein the freedom of recovering from applied strain via re-orientation and stretching of constituent entities is absent. So, although both polymers and colloidal suspensions deform under imposed stress or flow, but the latter has no strain-recovery mechanism. But colloidal suspensions may give rise to formation of local aggregates[14], slippage etc. as routes of viscous dissipation in response to the imposed stress.

These inter-particle interactions may be modulated via particle content of the colloidal suspension to display a range of complex rheological behaviour as a function of applied shear rates [15–18]. For example, dilute suspensions exhibit Newtonian flow behaviour but with an increase in particle weight fraction, ($\phi_w$), the suspension viscosity, ($\mu$) increases by several orders of magnitude. A continuous increase in suspension viscosity beyond a critical shear rate ($\dot{\gamma}_c$) marks the onset of continuous shear thickening (CST) [19–21]. Whereas, if the viscosity increase is abrupt, then, the behaviour is termed as discontinuous shear thickening (DST). Such transition from CST to DST with shear rate and particle content have been the subject of recent investigations [20,22–26].

In the present investigation, we experimentally probe the coalescence of one such model colloidal suspension, i.e., aqueous corn-starch suspensions[20,27–30]. These are also known to display an array of rheological responses and in particular concentrated corn-starch exhibit a transition from continuous shear thickening (CST) to dis-continuous shear thickening (DST) [20,27–35] with increase in shear rate. The question we probe is does this rheological behaviour influence the necking dynamics of coalescing colloidal droplets. Our results establish the connection between the rheological response of colloidal suspensions on the coalescence dynamics. Further, we attempt to employ the Cross model as the constitutive equation to account for the rheological behaviour of the suspensions and show that it captures the essential features of coalescence dynamics observed in the experiments.

## 2. MATERIALS & METHODS

**2.1. Substrate preparation.** Glass slides were procured (A – one, Haryana, India) and cleaned by ultrasonication in two solvents (acetone followed by deionized water) for 20 minutes each. These cleaned glass slides were dried in an oven at 90°C for 30 minutes to remove moisture. To have a sessile-pendant configuration, it was essential to have sessile droplets that exhibit partial wetting state with the substrate. Therefore, these substrates were coated with a hydrophobic layer of polydimethylsiloxane (PDMS). This hydrophobic layer was prepared by mixing the base with the curing agent (Sylgard 184, *Silicone Elastomer Kit, Dow Corning*) in the ratio of 10:1 and vacuum desiccated to remove the air bubbles. The mixture was then spin coated at 5000 rpm for 1 minute on the cleaned slides and cured in the hot air oven at 90°C for 90 minutes.

**2.2. Preparation of corn-starch suspensions.** Corn-starch powder (*Sigma-Aldrich*) was procured having average particle diameter, $d_{p(\mu m)}$ ~15.29 ± 4.37. Details of the particle size distribution are discussed in the **Supplementary information (Note S1)**. Corn-starch powder was added to aqueous solutions of cesium chloride (CsCl: water ~ 51.5:48.5 (w/w)) [32,33] to mitigate the settling of corn-starch particles. These mixtures were stirred using a magnetic stirrer at 650 RPM for 20 minutes to prepare the corn-suspensions of required weight fractions, $\phi_w$ (w/w). Aqueous solutions of cesium chloride ($\phi_w = 0$) are referred to as the control or reference suspensions in the text hereafter. The particle weight fractions were increased incrementally from typically dilute suspensions, $\phi_w = 0.26$ to concentrated suspensions, $\phi_w = 0.43$. Surface tension, ($\sigma$, mN/m) of these suspensions were measured by the pendant drop method using a goniometer (*Biolin Scientific,* Sweden, Theta lite) and found to lie in the range of 54 – 62 mN/m for the particle weight fractions employed in this study.

**2.3. Rheological characterization of cornstarch suspensions.** Corn-starch suspensions were characterized by measuring the variation of apparent viscosity ($\mu$) with imposed shear rate, $\dot{\gamma}$ ($10^{-2} - 10^3 \, s^{-1}$) using a rheometer (*Anton Paar*, model: MCR 302) with the cone and plate geometry (40 mm plate diameter, cone angle 1°).

With the introduction of colloidal particles, $\phi_w \sim 0.26$ the apparent viscosity showed a shear thinning regime at low shear rates followed by a Newtonian plateau at higher shear rate, $\dot{\gamma} > 1 s^{-1}$ (**Fig 1a**). Further increase in particle content showed the occurrence of a shear thickening behaviour at higher shear rates. To distinguish between the continuous and discontinuous shear thickening behaviour, the shear stress responses (**Fig.1b**) were fitted to $\tau \, \alpha \, \dot{\gamma}^{(\psi)}$ where the exponent, $\psi = 1$ corresponds to Newtonian regime. An increase in particle loading, $\phi_w$ i.e., $\phi_w > 0.26$, marked the onset of continuous shear thickening (CST) as evidenced by $\psi > 1$. With further increase in $\phi_w$ i.e., $\phi_w \sim 0.41$, an abrupt increase in apparent viscosity [30], indicated dis-continuous shear thickening (DST) behaviour with $\psi \sim 2$. Such dis-continuous shear thickening behaviour is typical of corn-starch suspensions and its causes are a still a topic of debate in the soft matter community [13,23,25,29,30,32,33,36–38]. This transition from CST to DST occurs at a particular shear rate, called critical shear rate, $\dot{\gamma}_c$.

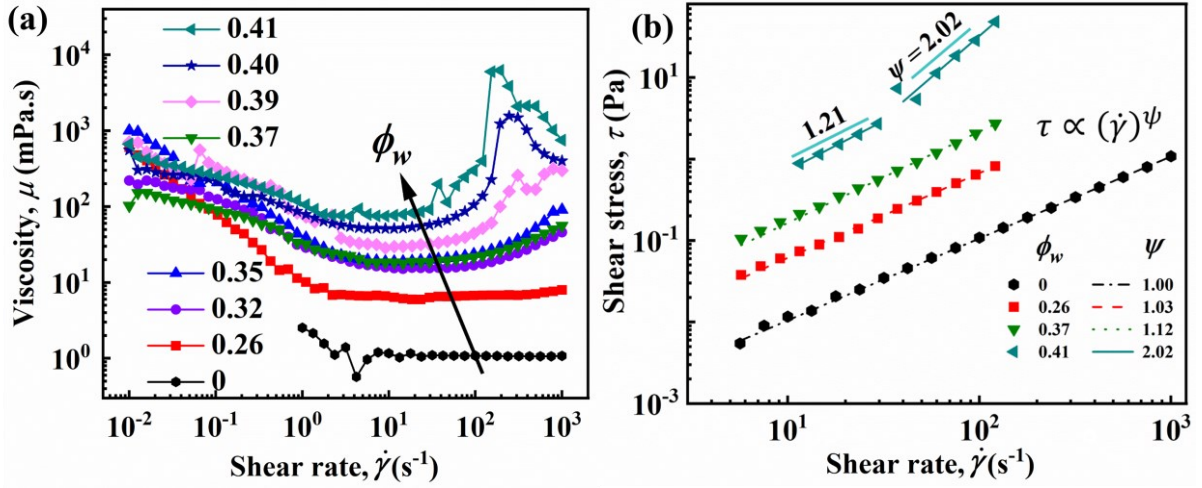

**Fig. 1** Rheological behaviour of corn-starch suspensions as a function of particle loading ($\phi_w$ : w/w ). Variation of **(a)** apparent viscosity ($\mu$) with applied shear rate ($\dot{\gamma}$). Arrow indicates the direction of increase in $\phi_w$ **(b)** shear stress ($\tau$) with shear rate ($\dot{\gamma}$) depicting the transition [30] of the rheological response from continuous shear thickening (CST) to discontinuous shear thickening (DST). The transition is determined by fitting the data to $\tau \propto \dot{\gamma}^{(\psi)}$ where $0 < \psi < 2$ and $\psi > 2$ indicates CST and DST resp. with increasing $\phi_w$. Fitted lines are shown for completeness.

The extensional behaviour of the colloidal suspensions are characterised by the capillary breakup extensional rheometry dripping on a substrate (CABER-DOS) experiments following the experimental protocol adapted from Dinic *et al*. [39]. This involves creation of a pendant drop by pumping the suspension through a nozzle (tip radius, $l_c$ = 0.6 mm) using a syringe pump operated at a constant flow rate of 0.02 mL/min. (*New Era Pump Systems Inc, USA*). The pendant drop is then slowly deposited on to a cleaned glass substrate and the capillary bridge formed between the colloidal suspension-substrate is captured by a high-speed camera (20,000 FPS) and compatible light source (*Nila Zaila, USA*). The gap, $H_d$ between the nozzle and the substrate, is kept constant, maintained at $H_d = 6l_c$. Details of the extensional behaviour of the suspension as a function of particle content are discussed in later sections.

**2.4. Coalescence experiments.** The sessile-pendant configuration is adopted to outline the effect of increasing particle content ($\phi_w$) on the coalescence dynamics of colloidal suspensions. Sessile droplets of constant volume (~ 7.5 μL) were dispensed onto the substrate. Pendant droplet of identical volume were brought in contact with the sessile drop to initiate coalescence at an approach velocity of $10^{-1}$ m/s [9,40] by a syringe pump (**Fig. 2a**). The temporal evolution of the neck formed between the coalescing droplets was captured (**Fig.2b, blow-up**) at 170,000 FPS using a high-speed camera (*Photron Fastcam Mini, AX - 100*) and a suitable light source (*Nila Zaila, USA*).

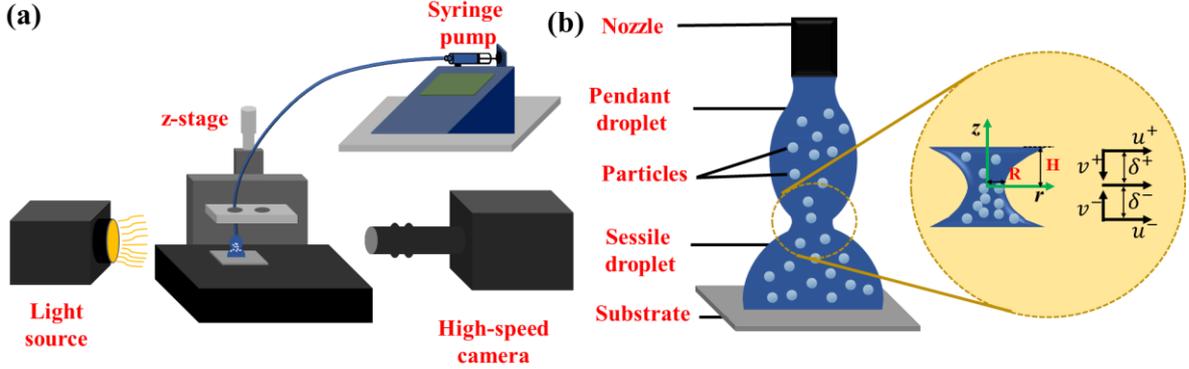

**Fig. 2** Schematic of the **(a)** experimental setup **(b)** coalescing sessile-pendant droplet. Blow-up of the region of interest (neck) shows experimental quantities of interest, neck radius, $R$ and semi-neck width, $H$ for colloidal suspension with particle weight fraction, $\phi_w$. Velocity field employed later in theoretical framework is shown in the blow-up.

Images were extracted from the captured real-time videos (**Supplementary video, S1**) of coalescence and the neck evolution was characterised by two parameters - radius, $R$ and semi-width, $H$ using an in-house MATLAB code with sub-pixel accuracy. Details of the analysis can be found elsewhere [7–10,40,41]. A blow-up of the coalescing droplets (**Fig.2(b)**) schematically shows velocity field in the neck region used for developing the theoretical formulation. Experiments were performed for each representative particle weight fraction and the average of at least five trials have been reported. Error range is found to lie within $\pm 5\%$ for the reported neck radii.

## 3. THEORETICAL FORMULATION

The flow in the neck region of coalescing droplets in the cylindrical co-ordinate system has been assumed to be incompressible, quasi-steady and quasi-radial, i.e., $V = (V_r, 0, V_z) := (u, 0, v)$ as shown in **Fig. 2(b)**. Considering the symmetry of the sessile-pendant geometry, the flow field near the neck region can be defined as, at $z = 0$, $u \neq 0$ and $v = 0$. In close proximity of the necking region, the flow velocity is given as, $u^+ = u^-$ and $v^+ = -v^-$.

The mass and momentum conservation equations at $z = 0$ can be expressed as eq. (1) and eq. (2) respectively,

$$\frac{\partial u}{\partial r} + \frac{u}{r} + \frac{\partial v}{\partial z} = 0 \quad (1) \qquad \rho\left(u\frac{\partial u}{\partial r}\right) = -\frac{\partial p}{\partial r} + \frac{\partial \tau_{rr}}{\partial r} + \frac{\tau_{rr}}{r} + \frac{\partial \tau_{rz}}{\partial z} \quad (2)$$

where $\tau$ is the stress tensor, $\tau = 2\eta(\dot{\gamma})\dot{D}$ with $\dot{D}$ as the strain rate tensor. These are further modified for droplets of a shear thickening fluid (STF) by introducing the Cross model [28,42–48]. The transition from CST to DST as a function of particle weight fraction is incorporated through four parameters, $\eta_\infty$, $\eta_o$, $\Gamma$ and $n$, $\eta(\dot{\gamma}) = \eta_\infty + \frac{\eta_o - \eta_\infty}{1+(\Gamma\dot{\gamma})^n}$ (3) where, $\eta_\infty$ and $\eta_o$ are the infinite and zero shear viscosity resp., $\Gamma$ is the time constant, $n$ is the power law index and $\dot{\gamma} = \sqrt{\dot{D}:\dot{D}} = \sqrt{\left(\frac{\partial u}{\partial r}\right)^2 + \left(\frac{\partial v}{\partial z}\right)^2 + \left(\frac{u}{r}\right)^2}$ is the second invariant of the strain rate tensor. The numerical values of these parameters were extracted by fitting the experimental data with eq. (3) (**Table S2, Supplementary information**). This model assumes the fluid viscosity to be a function of only the applied shear rate, $\dot{\gamma}$ in tune with rheological behaviour **(Fig. 1c)** of the suspensions. Modulation of the particle weight fraction is captured exclusively through the altered viscosity behaviour of the fluid and is incorporated through the individual stress tensor components along with the Cross constitutive relation, given by eq. (4),

$$\tau_{rr} \sim \eta_\infty \left(\frac{\partial u}{\partial r}\right) + \frac{\eta_o - \eta_\infty}{\Gamma^n}\left(\frac{\partial u}{\partial r}\right)^{1-n} \quad (4a) \qquad\qquad \tau_{rz} \sim 0 \quad (4b)$$

It must be noted that the term $1 + (\Gamma\dot{\gamma})^n$ has been approximated to $\sim (\Gamma\dot{\gamma})^n$ since, $O(\dot{\gamma}) \gg 1$. The *r-direction* momentum (eq. (2)) is further simplified by introducing the relevant scaling parameters, $u \sim U$, $r \sim R$, $z \sim \frac{R^2}{2R_o}$ while the pressure gradient scaled as, $\frac{\partial p}{\partial r} \sim \sigma\left(\frac{1}{R^2} + \frac{2R_o}{R^3}\right)$ where $\sigma$ is the measured surface tension of each suspension. The corresponding stress tensor components in terms of scaled variables can be written as, $\frac{\partial \tau_{rr}}{\partial r} \sim \frac{\tau_{rr}}{R}$ and $\frac{\partial \tau_{rz}}{\partial z} \sim \frac{\tau_{rz}}{\frac{R^2}{2R_o}}$ where, $R_o$ is the initial droplet radius The momentum conservation with these arguments is reduced to a polynomial equation with two scaling constants, $C_1, C_2$

$$U^2 + F(\Gamma,\rho,R,n,\eta_o,\eta_\infty)U^{1-n} + G(\rho,R,\eta_\infty)U + H(\sigma,\rho,R,R_o) = 0 \quad (5)$$

where, the scaling constants have been defined as,

$$F(\Gamma,\rho,R,n,\eta_o,\eta_\infty) = -2C_2\frac{\eta_o - \eta_\infty}{\rho\Gamma^n R^{1-n}} \quad (5a); \qquad G(\rho,R,\eta_\infty) = -2C_2\frac{\eta_\infty}{\rho R} \quad (5b);$$

$$H(\sigma,\rho,R,R_o) := -C_1\frac{\sigma}{\rho}\left(\frac{1}{R} + \frac{2R_o}{R^2}\right) \quad (5c)$$

Eq. (5) is solved, using iterative Newton Raphson method with initial guess corresponding to Newtonian fluid i.e., roots of eq. (5) for $n = 1$. Further, the evolution of the necking radius, $R$ is obtained by rewriting the velocity term, $U$ as, $U = \frac{dR}{dt}$ in eq. (5),

$$\frac{dR}{dt} = K(\sigma, \rho, \Gamma, R, R_o, n, \eta_o, \eta_\infty, C_1, C_2) \quad (6)$$

Solution of eq. (6) was obtained by the finite difference scheme with sufficiently small-time step ($10^{-6}$ s) to ensure the numerical stability of the scheme. The neck radius $R$ and time $t$ were non-dimensionalized by the length-scale corresponding to initial droplet radius, $R_o$ and the inertial time scale, $t_i = (\rho R_o^3/\sigma)^{0.5}$ respectively. These theoretically predicted temporal profiles of the necking radius are shown in the later sections.

## 4. RESULTS & DISCUSSIONS

The initial contact between the sessile and pendant droplets is a point that gives way to the development of a liquid bridge between the two droplets. In the present study, the lower limit of the region of interest (ROI) is the onset of liquid bridge formation where it can be captured (**Fig. 3(a)**) by the high-speed camera. Details of the methodology adopted for initiation point detection for coalescing colloidal droplets can be found in the **Supplementary information (Note S3).** The reader is also referred to a recent article [7] for details on discerning the ROI. Evolution of the liquid bridge is quantified by the neck radius, $R$ tracked temporally for the colloidal suspensions with incremental increase in particle weight fraction, $\phi_w$ i.e., from dilute to concentrated suspensions.

### 4.1 Colloidal droplet coalescence

Snapshots of coalescing droplets of solvent, $\phi_w = 0$ and colloidal suspension $\phi_w = 0.26$ are shown in parallel in **Fig. 3(a)**. It can be observed that the shape of the bridge for the representative colloidal suspension is similar to its suspending solvent. This implies that while initial bridge shape is unaltered by introduction of micron-sized colloids in droplets (dilute suspensions), its influence on neck growth dynamics is evident. For concentrated suspensions, like $\phi_w$ = 0.42 the initial neck radius is larger compared to the solvent and dilute suspension, $\phi_w$ = 0.26, but the neck growth occurs at a slower rate. The rate of neck growth within the ROI, is delayed with increasing particle content.

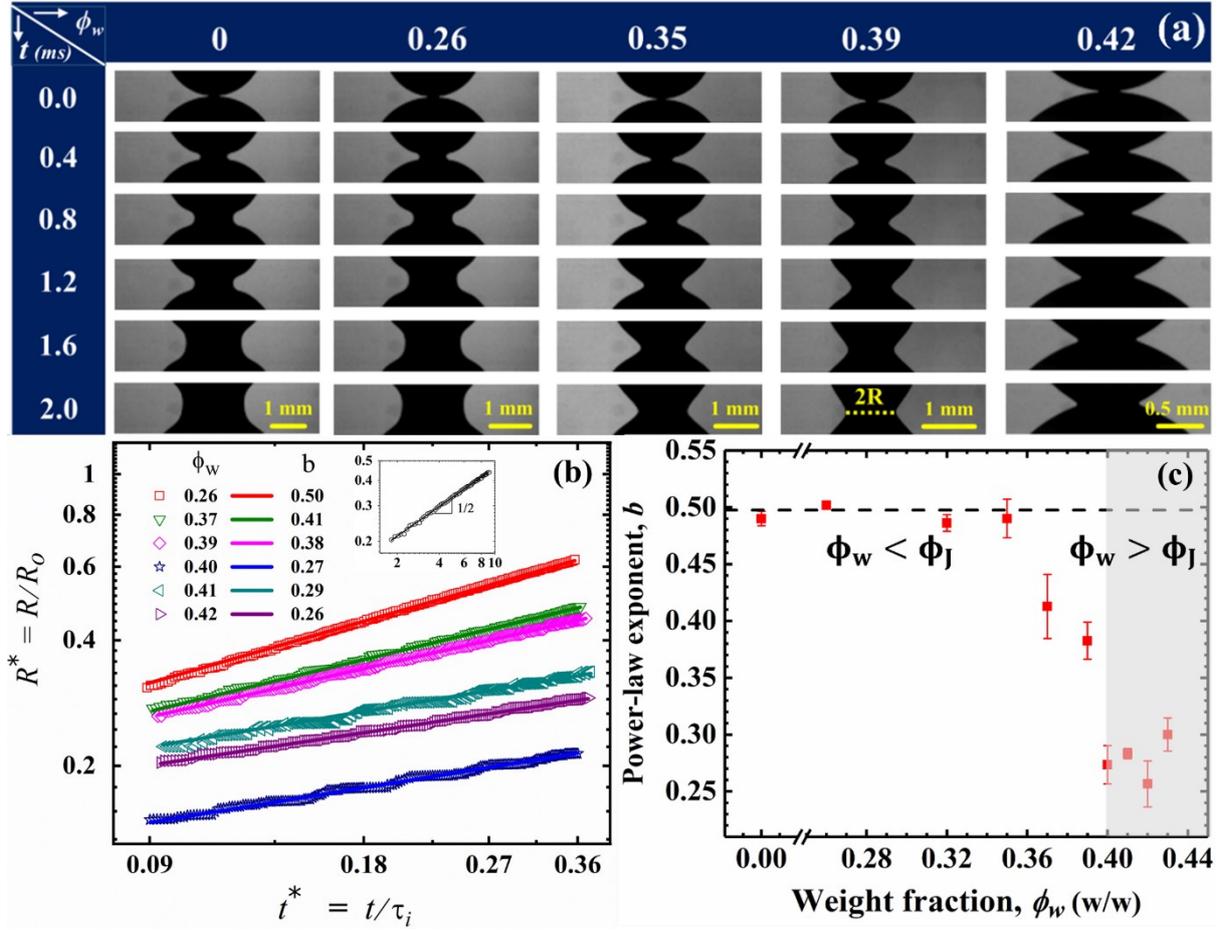

**Fig. 3** Temporal evolution of the neck radius, $R$ with increasing (**L to R**) particle weight fractions, $\phi_w$ for colloidal droplets shown via **(a)** real time images where, $R$ represents the neck radius during coalescence **(b)** corresponding power law behaviour $R = at^b$, where $a$: pre-factor and $b$: power-law exponent, given as slopes of the linear fits. Here, $R^*$ and $t^*$ are non-dimensionalised neck radius and time resp. $R_0$ is the initial droplet radius and $t_i (= \rho R_o^3/\sigma)^{0.5}$ is the inertial timescale ; $\sigma$ (mN/m) : suspension surface tension. Inset shows the typical $R = at^{0.5}$ for Newtonian suspension, $\phi_w = 0$ with $t^* \times 10^{-2}$ **(c)** variation of $b$ with particle weight fraction, $\phi_w$ where $\phi_J$ denotes particle jamming fraction.

This is further quantified in (**Fig. 3 (b)**) where the neck radius, $R$ and time are non-dimensionalised by the initial droplet radius, $R_0$ and inertial time scale, $t_i (= \rho R_o^3/\sigma)^{0.5}$ resp. with $\sigma$ (mN/m) is the corresponding measured suspension surface tension. The non-dimensionalised neck radius, $R^*$ and time, $t^*$ resp. exhibited power-law behaviour, $R = at^b$ with $b$ as the magnitude of the power-law exponent, (**Fig. 3(c)**). This power law exponent is found to be specific to each particle weight fraction ($\phi_w$). Specificity of the power-law exponent, $b$ to the particle weight fraction is further probed. In particular, does the magnitude of $b$ reflect the varying rheological characteristics (**Fig. 1**) of the colloidal suspensions observed with specific particle content.

Dilute colloidal suspensions, $\phi_w = 0.26, 0.32, 0.35$ exhibited coalescence dynamics ($b = 0.5$) similar to the suspending Newtonian solvent. But with an increase in particle loading from $\phi_w = 0.35 - 0.41$, a significant drop in the magnitude of $b$ (0.50 to 0.27) was observed. The steepest decrease in the power law exponent ($b = 0.38$ to $0.27$) occurs from increase in $\phi_w = 0.39$ to $0.40$. This coincides with the particle weight fraction, $\phi_w \sim 0.41$ where the CST to DST transition was observed to occur on account of imposed shear (**Fig. 2(b)**). $\phi_w \sim 0.41$ is also termed as '*dynamic jamming point*' ($\phi_J$) in the literature [24,26,27,31,49] where particles are known to self-aggregate forming clusters as a response to the imposed shear [32]. These clusters are known to dramatically alter the rheological response from fluid-like to solid-like with signs of pseudo elasticity. Evidence of the same can be seen in the oscillatory shear rheology response (**Fig. S3**) with $\frac{(G')}{(G'')} > 10$ at $\phi_w \sim 0.41$ for frequency $\omega > 5 \ (s^{-1})$. Beyond the jamming point $\phi_w > \phi_J$, the neck radius continues to exhibit power law behaviour with $b \sim 0.27$. However, the suspensions corresponding to $\phi_w = 0.42, 0.43$ consist of particle clusters regions disparately distributed throughout the solvent. These particle clusters give rise to localised density variations within the suspension [34]. It has been shown via rotational and capillary rheometry that the macroscopic response of DST by corn-starch suspensions stems from the separation of the colloidal system into regions of '*low-density-flowing*' and '*high-density-jammed regions*' [34]. An underlying implication is that global homogeneity of the suspension is no longer maintained. Taking the analogy further, it can be stated that during coalescence of concentrated suspensions for particle fractions beyond $\phi_J$, the flow within the neck is restructured. This restructuring however does not allow the suspension to behave like its Newtonian solvent for then the power law exponent would be its universal value of 0.5.

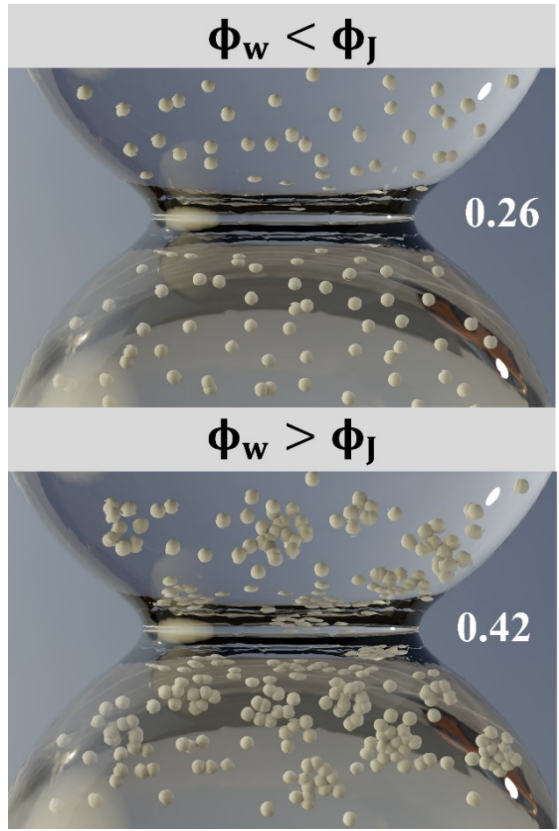

**Fig. 4**. Schematic representation of coalescing colloidal droplets for particle weight fraction beyond jamming point, $\phi_w > \phi_J$ where regions of particle clusters are formed.

Instead, it lies in the vicinity of $b \sim 0.27$, suggesting that the delayed neck growth dynamics comes from local flow restructuring and flow inhomogeneity.

Similar evidence of phase separation has been reported in Non-Brownian concentrated PVC suspensions [25]. Thus, the journey of the neck evolution can be divided into regimes based on the jamming point as shown schematically in **Fig. 4**. Also, we partially answer our initial query of rheological behaviour influencing coalescence dynamics of colloidal droplets in affirmation. Simply put, the rheological characteristic of the colloidal suspension has a bearing on its coalescence dynamics for $\phi_w > \phi_J$.

**4.2 Insights from pinch-off dynamics**

The correlation between the particle '*jamming*' referred to in the previous section is defined exclusively based on the response of colloidal suspensions subjected to shear flow [27,30,35]. But coalescence in the sessile-pendant droplet configuration is predominantly an extensional flow. Therefore, extensional behaviour of the colloidal suspensions was further probed via CABER-DoS experiments. Temporal evolution of the capillary bridge in these experiments were quantified by the non-dimensionalized capillary bridge radius $\bar{r}$ ($\bar{r} = r/l_c$, $l_c, r$ are the nozzle and capillary bridge radius resp.) as a function of the particle weight fraction (**Fig. 5**). Introduction of colloidal particles to the aqueous cesium chloride solution (**Fig. 5a, $\phi_w \sim 0$**) alters the characteristic shape of the capillary bridge. Further increase in particle content, $\phi_w \sim 0.32$ resulted in the appearance of the typical beads-on-a-string structure appears at the lowest particle weight fraction, **(Fig. 5a)** in contrast to the abrupt breakup seen in Newtonian solvent **(first row, Fig. 5a)**. The capillary bridge further stretches into a thin filament (filament thinning) thereby prolonging the pinch-off, at $\phi_w \sim 0.37$. This slow-down increases the time of pinch-off, $t_p$ [36,38,57,39,50–56] by an order of magnitude i.e., $\sim 10^{-2}$ at $\phi_w \sim 0.32$ to $\sim O(10^{-1})$ at $\phi_w \sim 0.4$ (**Fig. 5b**). The prolonged pinch-off dynamics occurs at the particle weight fraction range corresponding to the steepest decrease in power law exponent $b = 0.38$ to $0.27$ of coalescing colloidal droplets.

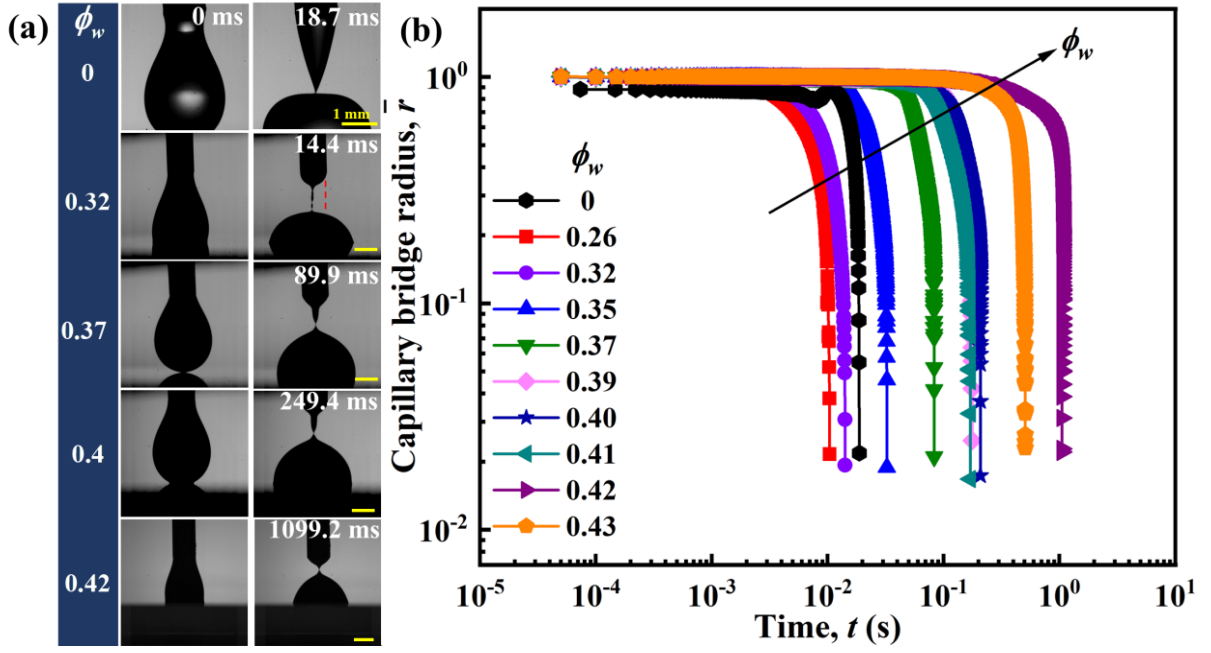

**Fig. 5 (a)** Real time snapshots show the prolonging of the capillary bridge evolution and eventual pinch-off with an increase in particle content, $\phi_w$. Left and right columns denote the capillary radius at t = 0 and before pinch-off resp. Dotted line emphasizes the filament thinning. **(b)** corresponding temporal evolution of the non-dimensionalised capillary bridge radius $\bar{r}$. Arrow indicates the direction of increase in $\phi_w$.

This can be attributed to the formation of local particle clusters and such particle confinement within the filament at the higher particle weight fractions ($\phi_w \geq 0.4$) are also known to spike the first normal stress difference [36,38,57,39,50–56]. The origin of the spike is particle confinement. As an analogy, such delay in pinch-off caused by filament thinning has also been reported in polymers; although the physical origins of filament thinning in polymers lies in the macromolecular re-orientations of polymer chains. Thus, filament thinning is a manifestation of polymer viscoelasticity. Similarly, the particle clusters generate the same global response of the colloidal suspensions under extensional flow i.e., signs of filament thinning or pseudo-elasticity. This hypothesis has been recently substantiated by experimental reports on capillary breakup dynamics of colloidal suspensions [38,54,57] with particle weight fractions in equivalent range as the present study.

The material response of colloidal suspensions [58–60] to extensional flow is fundamentally distinct from that seen in shear flow. However, there are only a handful studies [36,54,61] on extensional behaviour of corn-starch suspensions. To summarise those studies, the suspensions can be divided into less concentrated suspensions that exhibited a landscape of particle clusters giving rise to concentration in-homogeneities in the direction of pull (axial). Whereas, the relatively concentrated suspensions showed jamming [20,27,29–33,35,62]. It must be noted here that the assumption of a truly uni-axially extensional flow is questionable and suspension homogeneity often does not hold. Also, CaBER setup employed here in is essentially a *self-extensional* flow implying the fluid drainage by gravity will require infinite time for coalescence to be arrested.

**4.3 Colloidal droplet coalescence: subset of Sub-Newtonian coalescence**

To elucidate the mechanism responsible for neck growth modulation by particle weight fraction, a discussion of the forces at play in the system are considered hereafter. The neck growth is driven by the capillary force, $F_c$ and opposed by the inertia, $F_i$ and viscous forces, $F_v$ An additional force has to be taken into account for colloidal suspensions, i.e., the Brownian force, $F_b$. Thus, the expression for force balance in the coalescence phenomenon for colloidal droplets is, $F_c \sim F_v + F_i + F_b$ (7) The relative magnitudes of these forces can be estimated by the triad of time-averaged dimensionless numbers - Peclet number, $<Pe> = <\frac{\mu u_c d_p^3}{l_c kT}>$; particle Reynolds number, $<Re_p> = <\frac{a^2 \rho_f \dot{\gamma}_c}{\mu_c}>$; Stokes number, $Stk = <\frac{\rho u_c d_p^2}{18 l_c \mu}>$, quantified as a function of particle weight fraction, $\phi_w$ (**Fig. 6**). Here, $\rho, \mu_c, d_p (= 2a)$ denote the suspension properties i.e., density, viscosity at the critical shear rate ($\mu = \mu_c$) and the average particle diameter respectively, whereas $k$ and $T$ are the Boltzmann constant and absolute temperature (298 K). These dimensionless numbers quantify the governing forces at the characteristic length and velocity scales given by, $l_c \sim R$ and $u_c \sim \frac{\partial R}{\partial t}$. Note, the choice of viscosity magnitude $\mu$ is debatable since unlike Newtonian fluids, here viscosity is a function of applied shear rate for a specific particle fraction. The possible values include the zero-shear viscosity ($\mu_o$), viscosity at critical shear rate ($\mu_c$), and infinite shear viscosity ($\mu_\infty$). To capture the transition from CST to DST, the analyses here consider the viscosity at critical shear rate ($\mu = \mu_c; \dot{\gamma} = \dot{\gamma}_c$) for each suspension up to particle weight fractions of $\phi_w = 0.41$. This critical shear rate, $\dot{\gamma}_c$ marks the onset of shear thickening and is found to lie in the range of $\dot{\gamma}_c \sim 5$ to $25\ s^{-1}$ i.e., $O(10^0 - 10^1)$ for the investigated particle weight fractions.

The time-averaged Peclet number (**Fig. 6 (a)**) for the range of particle weight fraction investigated in this study is found to be greater than $O(10^5)$ implying the suspensions are non-Brownian. The effect of thermal forces are negligible, that reduces the force balance to, $F_c \sim F_v + F_i$ (8) The magnitude of Peclet number for a particular particle fraction as a function of imposed shear rate, $\dot{\gamma}$ ($Pe = 6\pi a^3 \mu \dot{\gamma}/kT$) has also been estimated (**Fig. 6 (b)**). Since, $\mu = f(\dot{\gamma})$, at low shear rates ($10^{-2} - 10^{-1}$), an increase in the particle weight fraction has negligible the effect on the Peclet number. But, at higher shear rate, i.e., $\dot{\gamma} > 10^1$ a jump of approximately four orders, from $O(10^5)$ to $O(10^9)$ is seen with the rise in particle content (**Fig. 6 (b)**). This range $O(Pe) \sim 10^4$ coincides with the order of Peclet number reported [33] for inception of shear thickening in identical system of colloidal suspensions (corn-starch suspensions). Further, it is found that the $Stk \sim O(10^{-3})$ implying imposed flow field timescale is greater than particle relaxation time scale. Thus, irrespective of the particle weight fraction, the particles are relaxed in the given flow field.

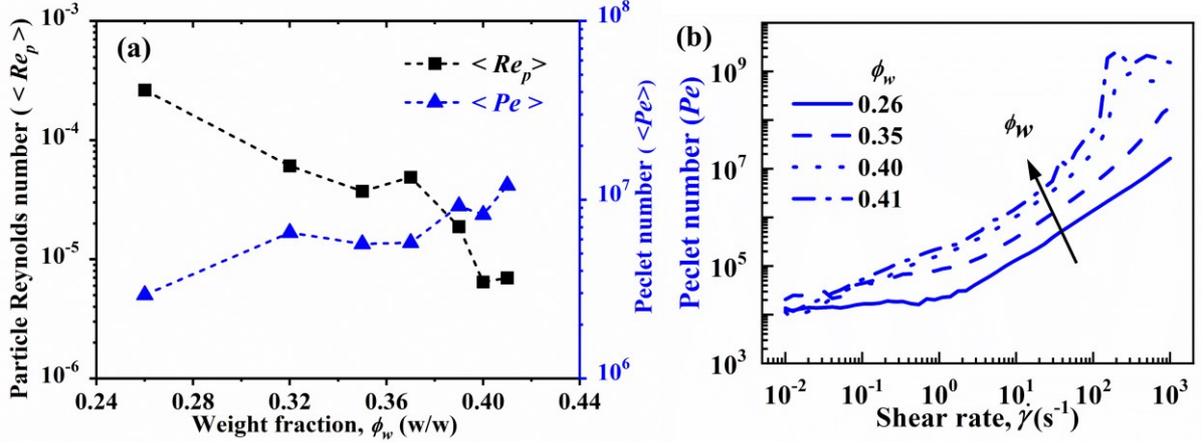

**Fig. 6 (a)** Variation of the time averaged dimensionless numbers for the system - particle Reynolds number ($< Re_p >$, ■) and Peclet number ($< Pe >$, ▲) as a function of particle weight fraction, $\phi_w$. Dotted lines are guide to reader's eyes only. **(b)** Variation of Peclet number, $Pe$ with imposed shear rate, ($\dot{\gamma}$) for representative particle weight fractions, $\phi_w \sim 0.26, 0.35, 0.4$ and $0.41$.

Further, the inertial effects can be neglected for analysis since, $Re_p \ll 1$. Thus, it can be stated that there is an absence of inertia-dominated regime. The deviation in the power law exponent with increase in particle content therefore arises from viscous effects as a consequence of the resistance faced by the particle clusters to pass through the small neck region with characteristic length-scale $\sim R \sim O(10^{-3})$. A deviation from the universal value of Newtonian fluid ($b \sim 0.5$) has been reported for polymeric fluid droplets, $b \leq 0.38$ [7,9,40] in the inertia-dominated regime. This reduction signifies the slow merger of polymeric droplets analogous to the sub-diffusive motion of Brownian particles in viscoelastic fluids. Taking a cue from the molecular world, such coalescence dynamics has been recently termed as Sub-Newtonian coalescence. The major contributor to the 'slowed' merging of polymeric fluid droplets is the additional degree of freedom of macromolecules i.e., its elasticity. This allows reorientation and stretching of macromolecular polymer chains. Similar re-arrangements of molecules can occur in a plethora of complex fluids ranging from magneto-rheological fluids (MRFs), electro-rheological fluids (ERFs) to liquid crystals under the action of pertinent external field. Extending the analogy further to colloidal suspensions, the particle-particle interactions start to dominate at higher shear rates for concentrated non-Brownian suspensions. This causes rupture of the solvent film around each particle and frictional contacts drive the formation of local particle aggregates. The formation of local particle clusters for $\phi_w \geq 0.35$ and their orientation along the direction of fluid flow in the neck-region acts as an additional opposing force. Thus, the reduction in the power law exponent $b < 0.5$ stems from hydrodynamic and frictional interactions between particle clusters rather than stretching or molecular reorientations. To summarize, coalescing colloidal droplets is a sub-set of the broad umbrella of sub-Newtonian coalescence.

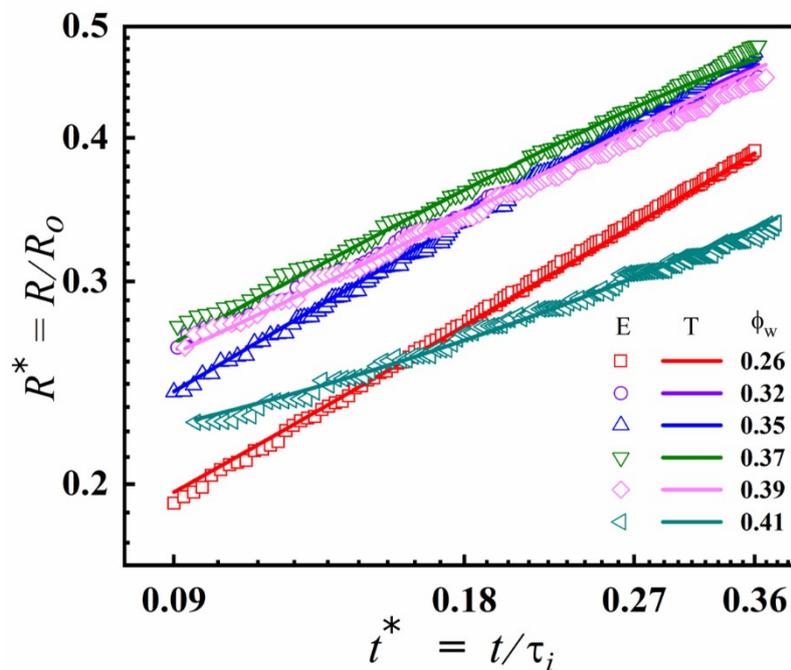

**Fig. 7** Superposition of theoretical temporal profiles ($T$) of the neck radius over the experimental profiles ($E$) as a function of the particle weight fraction, $\phi_w$. Here, $R^*$ and $t^*$ are non-dimensionalised neck radius and time resp. $R_0$ is the initial droplet radius and $t_i (= \rho R_o^3/\sigma)^{0.5}$ is the inertial timescale. $\sigma$ (mN/m) : suspension surface tension.

The role of inter-particle interactions in formation of particle clusters is accounted through the changes in suspension viscosity. This is captured by rheometry based characterisation of suspension rheology. A snowball effect of the rheological response can be seen in the phenomenological response, i.e., coalescence rate. Notably, at the jamming concentration, $\phi = \phi_J = 0.4$, a sharp decrease in the power law exponent of coalescence, $b$ can be observed. This transition is a manifestation of solid-like behaviour of the colloidal suspension due to the dispersed local particle clusters. The proposed model explains the behaviour through the scaling constants that are implicit functions of fluid viscosity parameters - $\mu, \mu_\infty, \mu_o$ etc. The mapping of predicted non-dimensionalised neck radius with the experimentally extracted data is shown in **Fig. 7**. It is evident that the proposed model predicts the experimental necking behaviour successfully. The only drawback of this model is it is applicable till $\phi = 0.41$. To include the higher particle weight fractions, $\phi =$ 0.41, 0.42, it would be necessary to quantify the distribution of particle-cluster-rich region and solvent-rich region in experiments and incorporating the same in the theoretical framework. This is beyond the scope of the present work and will be addressed in subsequent investigations.

The burgeoning interest in development of microfluidic devices [63–65] for DNA-based applications [66] and point of care diagnostics involve droplets with suspended micro/nanoparticles i.e., colloids as the operational entities. This led to the exploration of the four basic droplet operations – creation, splitting, merging (coalescence) and transport, that form the cornerstone of microfluidic devices [63]. However, the dynamics of colloidal droplet coalescence remains far from being understood. We believe the present study will offer new perspectives on this front and assist in predicting coalescence behaviour as a function of particle content.

## 5. CONCLUSIONS

This work, for the first time, establishes for the first time via extensive experiments the role of particle content in modulating coalescence dynamics of colloidal droplets. The model colloidal suspensions chosen in this study was aqueous corn starch suspensions that display shear thinning and discontinuous shear thickening characteristics in dilute and concentrated form resp. We report that this typical shear thickening behaviour has a bearing on the coalescence dynamics. In particular for particle weight fraction beyond the jamming fraction, i.e., ($\phi_w > \phi_J$ (= 0.4)) coalescence is delayed for concentrated suspensions. This allows us to conclude that the evolution of the liquid bridge in coalescing colloidal droplets is another example of the sub- Newtonian coalescence. We attribute this delay to local particle cluster formation as a response to applied shear during coalescence. A analysis of the relevant forces prevalent in the system – capillary, inertial, viscous and Brownian, we show that viscous forces are the dominant opposing force to the driving capillary forces. These viscous forces are accounted through our proposed theoretical model and show excellent agreement with the experimentally observed necking behaviour. We hope the present study will serve as a starting point in handling shear thickening concentrated colloidal droplets subjected to droplet operations such as coalescence. This will supplement the designing of digital microfluidic platforms and diagnostic devices that employ colloidal suspensions or entities heavily such as DNA assays, proteomics etc.

**Declaration of competing interests**

The authors declare that they have no known competing financial interests or personal relationships that could have appeared to influence the work reported in this paper.

**Acknowledgements**

UUG acknowledges generous support received from SERB POWER grant SPG/2021/003516. AK acknowledges support from SERB grant CRG/2022/005381. The authors would also like to acknowledge the contribution of Mr. Sreyajyoti Mondal, School of Biomedical Engineering, IIT (BHU) for assisting in sketching the schematic of proposed coalescence mechanism.